\documentclass[twocolumn,pra,showpacs,superscriptaddress]{revtex4}
\pdfoutput=1
\usepackage{setspace}
\usepackage{amsfonts,amsmath,amssymb}
\usepackage{txfonts}
\usepackage{graphicx, color}
\usepackage{hyperref}

\begin{document}

\title{Temporal coherence, anomalous moments, and pairing correlations in the classical-field description of a degenerate Bose gas}

\author{T. M. Wright}
\affiliation{Jack Dodd Centre for Quantum Technology, Department of Physics, University of Otago, PO Box 56, Dunedin, New Zealand}
\affiliation{ARC Centre of Excellence for Quantum-Atom Optics, School of Mathematics and Physics, University of Queensland, Brisbane, Queensland 4072, Australia}

\author{P. B. Blakie}
\affiliation{Jack Dodd Centre for Quantum Technology, Department of Physics, University of Otago, PO Box 56, Dunedin, New Zealand}

\author{R. J. Ballagh}
\affiliation{Jack Dodd Centre for Quantum Technology, Department of Physics, University of Otago, PO Box 56, Dunedin, New Zealand}

\begin{abstract}
The coherence properties of degenerate Bose gases have usually been expressed in terms of spatial correlation functions, neglecting the rich information encoded in their temporal behavior. In this paper we show, using a Hamiltonian classical-field formalism, that temporal correlations can be used to characterize familiar properties of a finite-temperature degenerate Bose gas. The temporal coherence of a Bose-Einstein condensate is limited only by the slow diffusion of its phase, and thus the presence of a condensate is indicated by a sharp feature in the temporal power spectrum of the field. We show that the condensate mode can be obtained by averaging the field for a short time in an appropriate phase-rotating frame, and that for a wide range of temperatures, the condensate obtained in this approach agrees well with that defined by the Penrose-Onsager criterion based on one-body (spatial) correlations. For time periods long compared to the phase diffusion time, the field will average to zero, as we would expect from the overall $\mathrm{U}(1)$ symmetry of the Hamiltonian. We identify the emergence of the first moment on short time scales with the concept of $\mathrm{U}(1)$ symmetry breaking that is central to traditional mean-field theories of Bose condensation.  We demonstrate that the short-time averaging procedure constitutes a \emph{general} analog of the `anomalous' averaging operation of symmetry-broken theories by calculating the anomalous thermal density of the field, which we find to have form and temperature dependence consistent with the results of mean-field theories.
\end{abstract}


\date{\today}

\maketitle
\section{Introduction}\label{sec:Introduction}
The precise experimental characterization of the properties of Bose-condensed gases has motivated the development of theoretical methodologies that can provide accurate and comprehensive descriptions of the condensed gas behavior. The fundamental theoretical framework is provided by many-body quantum field theory, but in general this becomes tractable only within approximation schemes, of which the most common are based around Bogoliubov's idea of representing the condensed atoms by a classical mean field.  In the very simplest form, this gives rise to the ubiquitous Gross-Pitaevskii equation, where the mean field is interpreted as the wavefunction of the condensate. The solution of the Gross-Pitaevskii equation has provided a useful first approximation to a wide range of equilibrium and dynamical phenomena, but the equation describes only the condensate, and neglects all spontaneous and incoherent processes. There are many situations where the condensate is accompanied by a component of thermal atoms which can have an important influence on the system properties and behavior, and the early mean-field treatments have been extended to give some level of description of the noncondensed atoms, by employing factorization approximations to the thermal component of the quantum field \cite{Griffin96,Burnett99,Hutchinson00}.  Such \emph{self-consistent mean-field} theories are built on the fictional \cite{Leggett01} but convenient and intuitively appealing assumption that Bose condensation breaks the $\mathrm{U}(1)$ phase symmetry of the underlying quantum field Hamiltonian, resulting in the appearance of \emph{anomalous moments} of the field: moments of the field operator [such as the mean field $\langle\hat{\Psi}(\mathbf{x})\rangle$] which are formally zero in a state of fixed particle number, but which acquire nonzero values in the symmetry-breaking approximation.  These treatments have provided an improved description of a range of equilibrium or near-equilibrium phenomena, but suffer from internal consistency problems, and have had limited success in describing the dynamics of the condensate at higher temperatures (see Ref.~\cite{Proukakis08} and references therein).  We note that many of the equilibrium predictions of the symmetry-breaking mean-field descriptions are regained in more careful, number-conserving approaches \cite{Gardiner97a, Castin98, Morgan00, Gardiner07}, however those methods have not provided a broadly tractable approach for dynamical or higher temperature systems. 

In recent years, a set of techniques has been developed that provides a unified nonperturbative description of both equilibrium and dynamical behavior of Bose gases for a temperature range from zero to close to the critical temperature. These so-called classical-field (or c-field) techniques \cite{Sinatra01, Blakie08, Goral01} have been used to provide a quantitative description of a number of key experimental results and regimes beyond mean-field theory (see \cite{Blakie08} for a summary of the broad range of recent applications).  While the treatment superficially resembles the zero-temperature Gross-Pitaevskii theory, the interpretation of the central object of the theory, the classical field $\psi(\mathbf{x})$, is very different: rather than the condensate wavefunction, it is an approximation to the Bose \emph{field operator}, and provides a means to evaluate quantum-mechanical correlation functions and their time development.  These correlation functions can be calculated by ensemble methods \cite{Norrie06a,Polkovnikov03b}, or for the case of equilibrium Bose-gas thermodynamics, by ergodic Hamiltonian methods \cite{Blakie05}. 

In this paper, we demonstrate that rich information is encoded in the temporal behavior of Hamiltonian classical-field trajectories.  Indeed in \cite{Wright08} we found that the temporal correlations of a classical field revealed a strong signature of a quasicondensate-like structure in a spatially disordered (vortex liquid) phase.  Here we consider the temporal correlations of a classical field containing a true condensate.  The phase of a condensate is \emph{by definition} \cite{Penrose56} rigid across the spatial extent of the condensed mode, and the only condensate-phase fluctuations are \emph{global} ones, which imply a diffusion of the phase over time (see Ref.~\cite{Sinatra08} and references therein).  This diffusion restores the $\mathrm{U}(1)$ phase symmetry of the system in the ergodic (microcanonical) density of the field, and anomalous moments such as $\langle \psi(\mathbf{x}) \rangle$ evaluated in this density therefore have vanishing values, consistent with the formal many-body theory for conserved particles.  However, the time scale of this phase diffusion is typically long compared with the correlation times of thermal modes in the field \cite{Sinatra08}, and we thus find that the condensate can be identified from the short-time average of the field in a frame phase-rotating uniformly at the underlying (mean) phase-rotation frequency of the condensate.  In this way phase-symmetry breaking emerges naturally from the Hamiltonian classical-field formalism.  We demonstrate that this averaging procedure constitutes a \emph{general} analog of the `anomalous' averaging operation of symmetry-broken theories \cite{Griffin96} by calculating the \emph{anomalous thermal density}, which characterizes pairing correlations in the noncondensed component of the field which are induced by the interacting condensate. 

This paper is organized as follows: In Sec.~\ref{sec:formalism} we give a brief outline of the equilibrium classical-field formalism we use here, and review its usual interpretation as a microcanonical formalism.  In Sec.~\ref{sec:temporal_coherence} we discuss the emergence of a mean (first moment) of the classical field, and make a quantitative comparison to the condensate defined by the Penrose-Onsager measure of one-body coherence.  In Sec.~\ref{sec:pairing_correlations}, we consider the anomalous second moments which comprise the classical-field \emph{pair matrix}, and construct the anomalous thermal density of the field. In Sec.~\ref{sec:conclusions} we summarize and present our conclusions. 
\section{Formalism}\label{sec:formalism}
\subsection{PGPE formalism}
The general formalism of (projected) classical-field methods has recently been reviewed at length in \cite{Blakie08}, but for the reader's convenience we will outline the projected Gross-Pitaevskii equation formalism we use in this work.  The dynamics we study are governed by the well-known classical-field Hamiltonian defined
\begin{equation}\label{eq:cfield_H}
	H_\mathrm{CF} = \int d\mathbf{x}\, \psi^*(\mathbf{x})\Big[H_\mathrm{sp} + \frac{U_0}{2}|\psi(\mathbf{x})|^2\Big]\psi(\mathbf{x}),
\end{equation}
where the single-particle Hamiltonian is
\begin{equation}
	H_\mathrm{sp} = \frac{-\hbar^2\nabla^2}{2m} + \frac{m}{2}\Big[\omega_r^2(x^2 + y^2) + \omega_z^2z^2\Big],
\end{equation}
and the interaction strength $U_0= 4\pi\hbar^2 a/m$ with $m$ the atomic mass and $a$ the $s$-wave scattering length.  The projected classical field is given by $\psi(\mathbf{x}) = \sum_{n\in\mathbf{L}}a_n Y_n(\mathbf{x})$, where the sum is over the \emph{finite} set of single-particle eigenmodes [$H_\mathrm{sp}Y_k(\mathbf{x})=\epsilon_kY_k(\mathbf{x})$] with eigenvalues $\epsilon_n \leq E_R$, where $E_R$ is the single-particle \emph{cutoff energy}.  Defining the projector
\begin{equation}\label{eq:Pdef} 
	{\cal P}f(\mathbf{x})\equiv\sum_{n \in \mathbf{L}}Y_n(\mathbf{x})\int d\mathbf{y}\; Y_n^*(\mathbf{y})f(\mathbf{y}),  
\end{equation} 
we can express the Hamilton's equation for $\psi(\mathbf{x})$ obtained from Eq.~(\ref{eq:cfield_H}) as
\begin{equation}\label{eq:PGPE} 
	i\hbar\frac{\partial\psi(\mathbf{x})}{\partial t} = {\cal P}\left\{\left(H_{\mathrm{sp}}+U_0|\psi(\mathbf{x})|^2\right)\psi(\mathbf{x})\right\},
\end{equation} 
which is the projected Gross-Pitaevskii equation \cite{Blakie08}. The Hamiltonian $H_\mathrm{CF}$ is invariant under the $\mathrm{U}(1)$ (gauge) transformation $\psi(\mathbf{x})\rightarrow\psi(\mathbf{x})e^{i\theta}$ and has no explicit time dependence, so that the evolution described by Eq.~(\ref{eq:PGPE}) conserves both the normalization $N_\mathrm{c}[\psi]=\int d\mathbf{x} |\psi(\mathbf{x})|^2$ of the classical field and the classical-field energy defined by $H_\mathrm{CF}$.  In the microcanonical approach of the PGPE we follow here, finite-temperature equilibrium configurations of the classical field are obtained by evolving in real time randomized initial configurations constructed with a particular energy $E[\psi]=H_\mathrm{CF}[\psi]$, such that the field naturally approaches thermal equilibrium, due to the ergodic nature \cite{Lebowitz73} of the classical-field system. 
\subsection{System parameters}
In the remainder of this paper we will specify quantities in the characteristic units of the radial trapping potential, quoting frequencies in units of $\omega_r$, distances in units of $r_0=\sqrt{\hbar/m\omega_r}$, times in units of $\omega_r^{-1}$, and energies in units of $\hbar\omega_r$.  We consider a system with $\omega_z=\sqrt{8}\omega_r$ (representing a typical three-dimensional trap geometry), and interaction strength $N_\mathrm{c}U_0 = \sqrt{2}\times500\hbar\omega_r/r_0^3$.  The corresponding ground (Gross-Pitaevskii) eigenstate of the system has energy $E\approx9N_\mathrm{c}\hbar\omega_r$, and we choose the cutoff $E_R=31\hbar\omega_r$.  We form random initial states \cite{Blakie05,Blakie08} with energies in the range $E\in[9.5,24.0]N_\mathrm{c}\hbar\omega_r$, which we allow to equilibrate by evolving them in real time for a period of $120\omega_r^{-1}$, and perform our analysis on their subsequent evolution. 
\subsection{Microcanonical interpretation}
Here we briefly remind the reader of the microcanonical (ergodic) interpretation of the PGPE applied to equilibrium systems \cite{Davis01,Blakie05,Blakie08}.  The method exploits the (empirical) fact that the PGPE trajectories are \emph{ergodic}, and thus provide a sampling of the \emph{microcanonical density}
\begin{equation}\label{eq:mu_density}
	P[\psi;E] = \left\{
	\begin{array}{rl}
		\mathrm{const} & H_\mathrm{CF}[\psi] = E \\
		0 & H_\mathrm{CF}[\psi] \neq E .  
	\end{array} \right.  
\end{equation}
The trajectories $\psi(\mathbf{x},t)$ cover the density $P[\psi;E]$ densely, and so averages in the density $P[\psi;E]$ are increasingly well-approximated by time-averages along trajectories $\psi(\mathbf{x},t)$ of increasing length.  The implications of this for PGPE simulations are two-fold: First, a theorem due to Rugh \cite{Rugh97} shows that the temperature of a microcanonical system can be expressed as an average in its microcanonical density, and thus calculated from a time average.  Second, equilibrium \emph{correlation functions} of the classical field can similarly be defined as averages in the density (\ref{eq:mu_density}), and thus evaluated from time averages.

A correlation function of particular interest for characterizing condensation in the classical field is the covariance matrix defined
\begin{eqnarray}\label{eq:density_mtx}
	\rho(\mathbf{x},\mathbf{x'}) &\equiv& \langle \psi^*(\mathbf{x})\psi(\mathbf{x'}) \rangle_\mu \nonumber \\
	&=& \sum_j n_j \chi_j^*(\mathbf{x})\chi_j(\mathbf{x}'),
\end{eqnarray}
[where $\langle \cdots \rangle_\mu$ denotes a microcanonical average, i.e. an average in the ensemble with density given by Eq.~(\ref{eq:mu_density})], which forms the classical-field analog of the \emph{one-body density matrix}.  In the second line we have used the fact that $\rho(\mathbf{x},\mathbf{x}')$ is Hermitian to express it in a diagonalized form, where the coefficients $\{n_i\}$, indexed in order of decreasing magnitude, are the occupations of the corresponding modes $\{\chi_i(\mathbf{x})\}$.  By analogy to the criterion of Penrose and Onsager (PO) \cite{Penrose56}, condensation in the field is signaled by the most highly occupied mode $\chi_0(\mathbf{x})$ having an occupation $n_0$ which is significantly larger than all other occupations $n_i$.  This definition in terms of correlations in the microcanonical density is an unambiguous measure of condensation in the simple equilibrium regimes in which it is applicable.  Generalizations of this procedure based on \emph{short-time} fluctuation statistics have been applied to more general scenarios, involving (e.g.) broken rotational symmetries and nonequilibrium fields \cite{Blakie05,Bradley08,Wright09a,Wright10a}.  In the remainder of this paper we will simply refer to the classical-field covariance matrix Eq.~(\ref{eq:density_mtx}) as the one-body density matrix, and to the identification of its most highly occupied mode as the condensate as the \emph{PO approach} to quantifying condensation in the field.

It is important to note that the microcanonical density [Eq.~(\ref{eq:mu_density})] inherits the invariance under gauge transformations $\psi(\mathbf{x})\rightarrow\psi(\mathbf{x})e^{i\theta}$ of the Hamiltonian Eq.~(\ref{eq:cfield_H}).  Consequently only the averages of quantities which are invariant under such transformations are nonzero in the microcanonical density, which correspond of course to averages of operators which conserve particle number in the corresponding second-quantized field theory \cite{Blaizot86}.  Sinatra and Castin \cite{Sinatra08} have shown in a homogeneous geometry (where the condensate mode is \emph{a priori} the $k=0$ plane-wave state), that the classical-field condensate undergoes a slow phase diffusion.  This diffusion ensures that the gauge symmetry is restored in the microcanonical density.  In this paper, we show that on time scales short compared to the characteristic time scale of phase diffusion, the condensate is resolvable as the \emph{mean} of the field in an appropriate frequency-shifted frame. Furthermore, we show that higher anomalous moments can be similarly defined in terms of short-time averages of fluctuations about this mean field.
\section{Temporal coherence}\label{sec:temporal_coherence}
\subsection{Temporal coherence and the emergence of a nonzero first moment}\label{subsec:first_moment}
\subsubsection{Identification of the first moment}\label{subsubsec:id_first_moment}
We begin by quantifying the coherence of the time-dependent field $\psi(\mathbf{x},t)$ via its temporal power spectrum, evaluated at different spatial locations $\mathbf{x}$ \cite{Wright08}. We define the temporal power spectrum of the classical field $\psi$ at position $\mathbf{x}$, evaluated over a period of length $T$ as 
\begin{equation}\label{eq:power_spec}
	H(\mathbf{x};\Omega) = |\mathfrak{F}^T\{\psi(\mathbf{x},t)\}|^2,
\end{equation}
where $\mathfrak{F}^T\{f(t)\}$ denotes the Fourier coefficient taken from some arbitrary time origin
\begin{equation}\label{eq:four_coeff}
	\mathfrak{F}^T\{f(t)\} \equiv \frac{1}{T}\int_0^Tf(t)e^{i\Omega t}dt.
\end{equation}

\begin{figure}
	\includegraphics[width=0.50\textwidth]{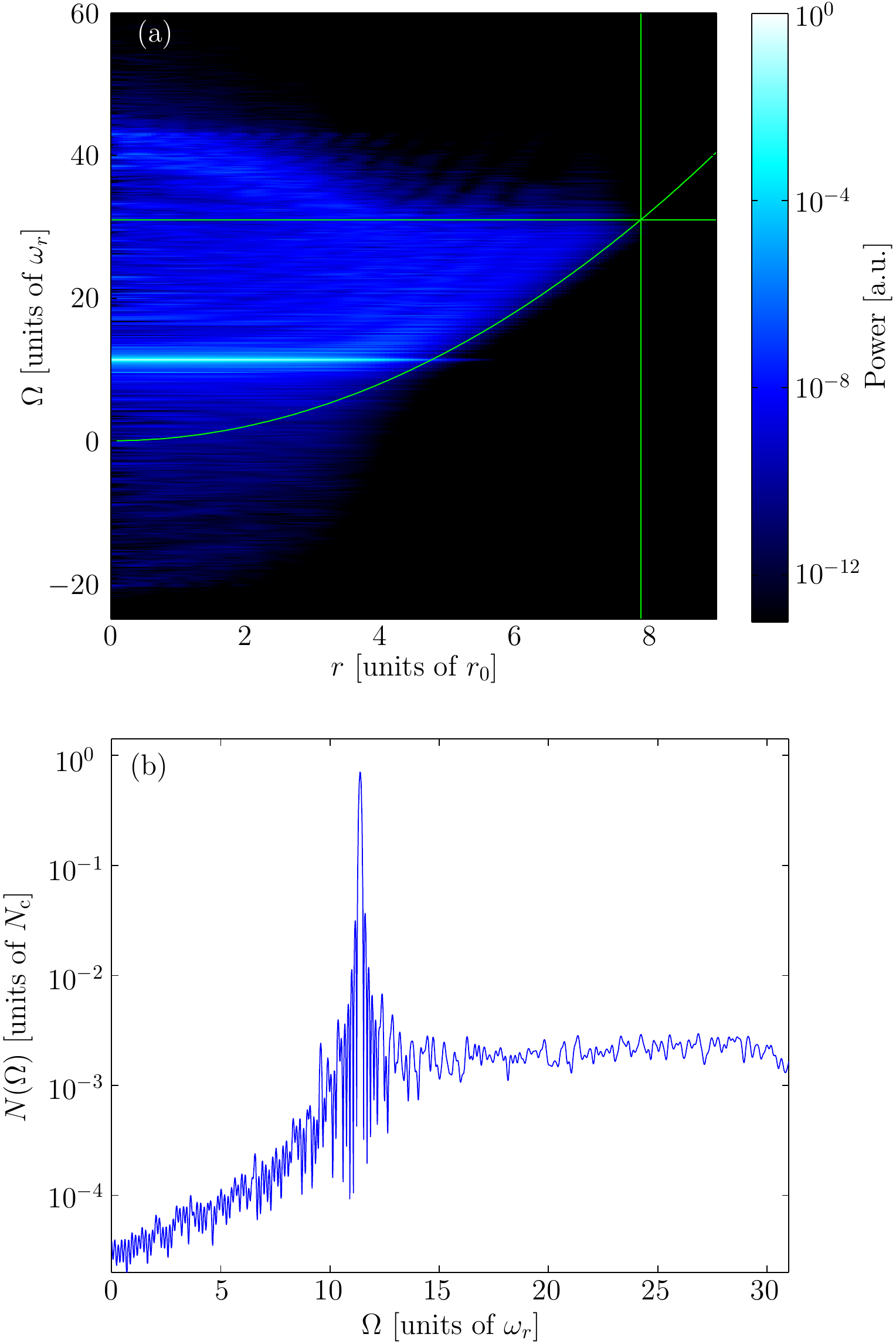}
	\caption{\label{fig:power_spectrum_in_trap} (Color online) (a) Power spectrum $H(\mathbf{x};\Omega)$ of the classical field on the plane $z=0$ (azimuthally averaged). Green (gray) lines indicate the trapping potential, cutoff energy and corresponding classical turning point of the trap. (b)  Space-integrated power spectrum $N(\Omega)$ of the field [see Eq.~(\ref{eq:N_omega})] as a function of the phase-rotation frequency $\Omega$.  Parameters of the classical field are given in the text.}
\end{figure}

In \cite{Wright08} we applied this procedure to a classical field in a disordered vortex-liquid state, in which spatial order of the system was strongly suppressed, and found a narrow peak in the power spectrum.  The appearance of such a peak is consistent with analytical results obtained by Graham \cite{Graham02} which suggest that \mbox{(quasi-)long}-range spatial order of the Bose field is accompanied by \mbox{(quasi-)long}-range \emph{temporal} correlations which decay in a functionally equivalent way.  Here we calculate the power spectrum for a classical field with the trapping and interaction parameters of Sec.~\ref{sec:formalism}, and energy $E=12.0N_\mathrm{c}\hbar\omega_r$.  Using the PO approach, we find that this field exhibits a (true) condensate, with condensate fraction $f_\mathrm{c}\equiv n_0/N_\mathrm{c}=0.70$.  We choose a sampling period of $40\omega_r^{-1}$, and approximate the integral in Eq.~(\ref{eq:four_coeff}) by a discrete sum over $1000$ equally spaced samples of the classical field.  In practice, we calculate the power spectrum at points in the $z=0$ plane, and average it over the azimuthal angle in this plane to smooth out fluctuations.  We thus obtain the averaged power spectrum as a function of the radius $r$, which we present in Fig.~\ref{fig:power_spectrum_in_trap}(a).  The oscillation frequencies $\Omega$ we measure in the time-dependent field correspond, of course, to energies $\epsilon=\hbar\Omega$ in the quantum mechanical system.   

For comparison, on the same figure we also plot the profile of the harmonic trapping potential $V(r)/\hbar$ in this plane [parabolic green (gray) line] and the cutoff energy $E_R/\hbar$ (horizontal line), and the classical turning point (vertical line) of the low-energy region $\mathbf{L}$ defined by their intersection.  The most prominent feature in this plot is the strong peak in the power spectrum centered on $\Omega=11.4\omega_r$, which is a signature of the long-lived temporal phase coherence in the classical field.  We identify the frequency $\lambda_0$ of this peak as the \emph{condensate frequency}.  The broad, lower intensity background spectrum represents the thermally occupied excitations in the classical-field system.  It is worth pointing out that in the central region of the trap ($r\lesssim 4r_0$), the background spectrum is strongly distorted by the presence of the condensate, with positive frequency components extending to approximately $\lambda_0+E_R/\hbar$, and negative frequency components appearing with energies extending down to approximately $\lambda_0-E_R/\hbar$ \cite{NoteA}.  Similar behavior was observed in Ref.~\cite{Wright08}, and represents the restructuring of the excitation spectrum of the trap by an interacting (quasi-)condensate, which distorts the single-particle excitations of the system into the familiar Bogoliubov particle-hole pairs \cite{Fetter99,Castin98,Mora03}. 

The temporal coherence we observe results from the quasi\-uniform phase rotation of the condensate:  the phase of the condensate exhibits a uniform rotation at frequency $\lambda_0$, superposed with a slow diffusion.  The width of the power spectrum peak here is thus determined by the rate of this global condensate-phase diffusion.  On time scales short compared with the characteristic time scale of phase diffusion, the condensate has an approximately constant phase in a frame co-rotating at frequency $\lambda_0$, i.e., short-time averages in this frame yield a nonzero \emph{first moment} $\langle \psi \rangle$ of the classical field.  A key observation of this paper is that time averages constructed in this way are analogous to the \emph{anomalous averages} which arise in symmetry-broken descriptions of Bose condensation \cite{Griffin96}, where the appearance of nonzero values for expectations of non-gauge-invariant quantities (i.e., the breaking of the phase symmetry) signals the presence of condensation in the field.  We thus consider the classical field frequency-shifted by $\Omega$ 
\begin{equation}
	\tilde{\psi}(\mathbf{x},t;\Omega) = e^{i\Omega t} \psi(\mathbf{x},t),
\end{equation}
and consider time-averages of this quantity formed from the same set of samples used to construct the power spectrum in Fig.~\ref{fig:power_spectrum_in_trap}(a).  We define the time-averaged field 
\begin{eqnarray}\label{eq:time_avgd_field}
	 \phi(\mathbf{x};\Omega) &\equiv& \langle \tilde{\psi}(\mathbf{x},t;\Omega) \rangle_t \\
	\Big(\! &=& \mathfrak{F}^T\{\psi(\mathbf{x},t)\} \;\Big), \nonumber
\end{eqnarray} 
where $\langle \cdots \rangle_t$ denotes a time average over a given period $T$ ($40\omega_r^{-1}$ in this case).  The time-averaged field $\phi(\mathbf{x};\Omega)$ is therefore the component of the classical field which phase-rotates like $e^{-i\Omega t}$, and its norm square quantifies the total (i.e., space-integrated) power contained in the field at frequency $\Omega$, i.e., 
\begin{equation}\label{eq:N_omega}
	N(\Omega) \equiv \int \! d\mathbf{x} \;|\phi(\mathbf{x};\Omega)|^2 = \int\! d\mathbf{x}\; H(\mathbf{x};\Omega).
\end{equation}
In Fig.~\ref{fig:power_spectrum_in_trap}(b) we plot this power as a function of the frequency $\Omega$, and note that it exhibits a prominent peak at $\Omega=11.38\omega_r$.  We identify the frequency at which the norm square of the time-averaged field (equivalently, the space-integrated power of the field) is maximized as the condensate frequency $\lambda_0$, and the corresponding time-averaged field $\phi(\mathbf{x};\lambda_0)$ as the classical-field condensate or \emph{mean field} \cite{NoteB}.  A nonzero time-averaged field occurs because the condensate has a reasonably well-defined phase on short time periods.  We identify this quasi-definite phase as an analog of the condensate phase which emerges in symmetry-broken descriptions of Bose-Einstein condensation; in this view point, the first moment $\phi(\mathbf{x};\lambda_0)$ is the analog of the condensate wavefunction $\langle \hat{\Psi}(\mathbf{x}) \rangle$ in such mean-field theories of Bose condensation.  For notational convenience, we introduce the norm square of the mean field $N_0\equiv N(\lambda_0)$, and the normalized mean-field mode function $\phi_0(\mathbf{x})\equiv\phi(\mathbf{x};\lambda_0) / \sqrt{N_0}$.  The norm square $N_0$ corresponds to the \emph{population} of the mean-field condensate mode, and we indeed find $N_0/N_\mathrm{c} = 0.706$, in close agreement with the PO value for the condensate fraction ($f_\mathrm{c}=0.70$).  To further compare this temporal-coherence method of identifying the condensate with the PO approach, we calculate the overlap of $\phi_0(\mathbf{x})$ with the eigenvector $\chi_0(\mathbf{x})$ obtained by the PO procedure.  We find $1- |\langle \phi_0 | \chi_0 \rangle| \approx 1.4\times 10^{-4}$, i.e., the condensate orbitals obtained by the two different procedures agree to a very high accuracy.  
\subsubsection{Temporal coherence and sample length}\label{subsubsec:temporal_coherence}
\begin{figure}
	\includegraphics[width=0.45\textwidth]{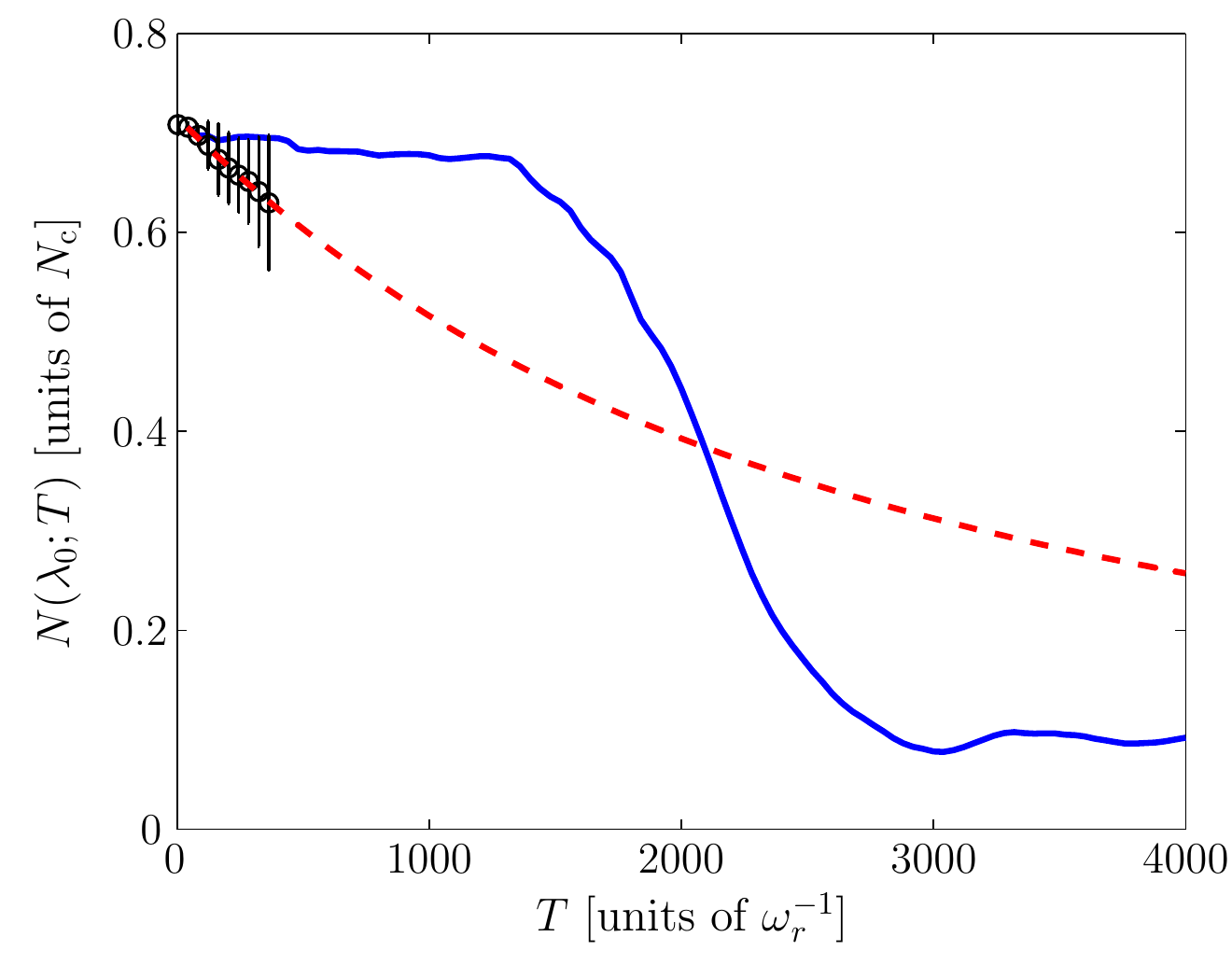}
	\caption{\label{fig:decay} (Color online) Norm square $N_0$ of the time-averaged field (i.e., space-integrated power of the classical field measured at the condensate frequency) as a function of the sampling period $T$. The solid line shows the value obtained from a single contiguous sampling of the classical field over period $T$.  Circles (with error bars) show the mean (and standard deviation) of estimates $N_0$ measured on $10$ individual $400\omega_r^{-1}$ sub-periods of the $4000\omega_r^{-1}$ time series.  The dashed line shows the expected (ensemble averaged) power, extrapolated from a least-squares fit to the means of the short-time estimates.  Parameters of the classical field are given in the text.}
\end{figure}

The results obtained for the mean field have an important dependence on the averaging time.  As discussed by Sinatra and Castin, the condensate phase exhibits diffusive evolution with time in the classical microcanonical ensemble \cite{Sinatra08}.  Consequently, we expect the power in the classical field measured at the condensate frequency to decay with time, exhibiting a power-law tail $N(\lambda_0;T) \sim 2/\gamma T$ at long times, as discussed in Appendix~\ref{app:AppendixA}.  We illustrate this issue using the same simulation ($E=12N_\mathrm{c}\hbar\omega_r$) as in the previous section.  Increasing the sampling period to $T\gtrsim50\omega_r^{-1}$ the condensate frequency is more accurately resolved as $\lambda_0=11.39\omega_r^{-1}$.  We assume this value as a best estimate for the condensate frequency, and calculate the power at this frequency as a function of the measurement period $T$, up to a maximum measurement period of $4000\omega_r^{-1}$.  In Fig.~\ref{fig:decay} we plot the power measured at frequency $\lambda_0$ (solid line), and find that it decays in a nonuniform way with increasing $T$.  However, the (normalized) mean-field orbital $\phi_0(\mathbf{x})$ we obtain at frequency $\lambda_0$ satisfies $1-|\langle\phi_0|\chi_0\rangle| \lesssim 10^{-4}$ for all averaging periods $T$ we consider; i.e., although the measured \emph{occupation} of the condensate decays with increasing averaging period due to the diffusion of the condensate phase, the \emph{mode shape} we obtain is relatively unaffected. 
The nonuniform decay of the mean-field orbital's occupation we observe is to be expected for a single trajectory, whereas we expect the scaling $N \sim 2/\gamma T$ to emerge from an average over a large ensemble of similarly prepared classical-field trajectories (cf. \cite{Sinatra08}).  It is possible, however, to infer $\gamma$ from a single trajectory, as we now show.  We divide the total $4000\omega_r^{-1}$ ($10^5$-sample) period of the classical-field trajectory into $10$ consecutive sub-periods of length $400\omega_r^{-1}$ (each of $10^4$ samples), and regard these sub-periods as an ensemble of 10 distinct trajectories.  For each member of the ensemble we calculate the power $N(\lambda_0;T)$ as a function of $T \leq 400\omega_r^{-1}$.  We then average over these 10 ensemble members to obtain a mean power estimate for each sampling period $T$.  The means and standard deviations of these measurements are indicated by circles with error bars in Fig.~\ref{fig:decay}, and by performing a least-squares fit of the expected power $\langle N(\lambda_0;T) \rangle$ at the condensate frequency [Eq.~(\ref{eq:mean_decay}) in Appendix~\ref{app:AppendixA}] to these mean power estimates, we estimate the phase-diffusion coefficient $\gamma\approx10^{-4}\omega_r$ \cite{NoteC}. The dashed line in Fig.~\ref{fig:decay} extrapolates the expected behavior of $\langle N(\lambda_0;T) \rangle$ to later times.  Given this decay of the peak power with $T$, a rigorous estimate of the condensate population would in principle be obtained by forming estimates $\langle N(\lambda_0,T_i) \rangle$ for multiple sampling period lengths $T_i$, and extrapolating the resulting trend back to $T=0$ to estimate the `true' condensate population.   However, due to the weak linear decay of the power spectrum peak at short sampling periods, we can accurately estimate the condensate population as the magnitude of the dominant peak in the power spectrum obtained over a short sampling period, for all but the smallest condensate fractions (see Sec.~\ref{subsubsec:phase_freq}).
\subsection{Dependence of the first moment on the field energy}\label{subsec:energy_dependence}
In the ergodic classical-field (PGPE) method, equilibrium field configurations of different temperatures can be formed simply by varying the (conserved) energy of the random initial field configuration \cite{Davis01}.  In this section we investigate the behavior of the first moment introduced in Sec.~\ref{subsec:first_moment} as the energy (and thus temperature) of the classical-field equilibrium is varied, and compare its mode shape $\phi_0(\mathbf{x})$ and occupation $N_0$ with the Penrose-Onsager condensate orbital $\chi_0(\mathbf{x})$ and occupation $n_0$, respectively.  We further compare the condensate frequency $\lambda_0$ to the microcanonical chemical potential $\mu$ of the field obtained using the methodology of \cite{Rugh97,Davis03,Davis05}.
\subsubsection{Condensate fraction}\label{subsec:cfrac}
\begin{figure}
	\includegraphics[width=0.45\textwidth]{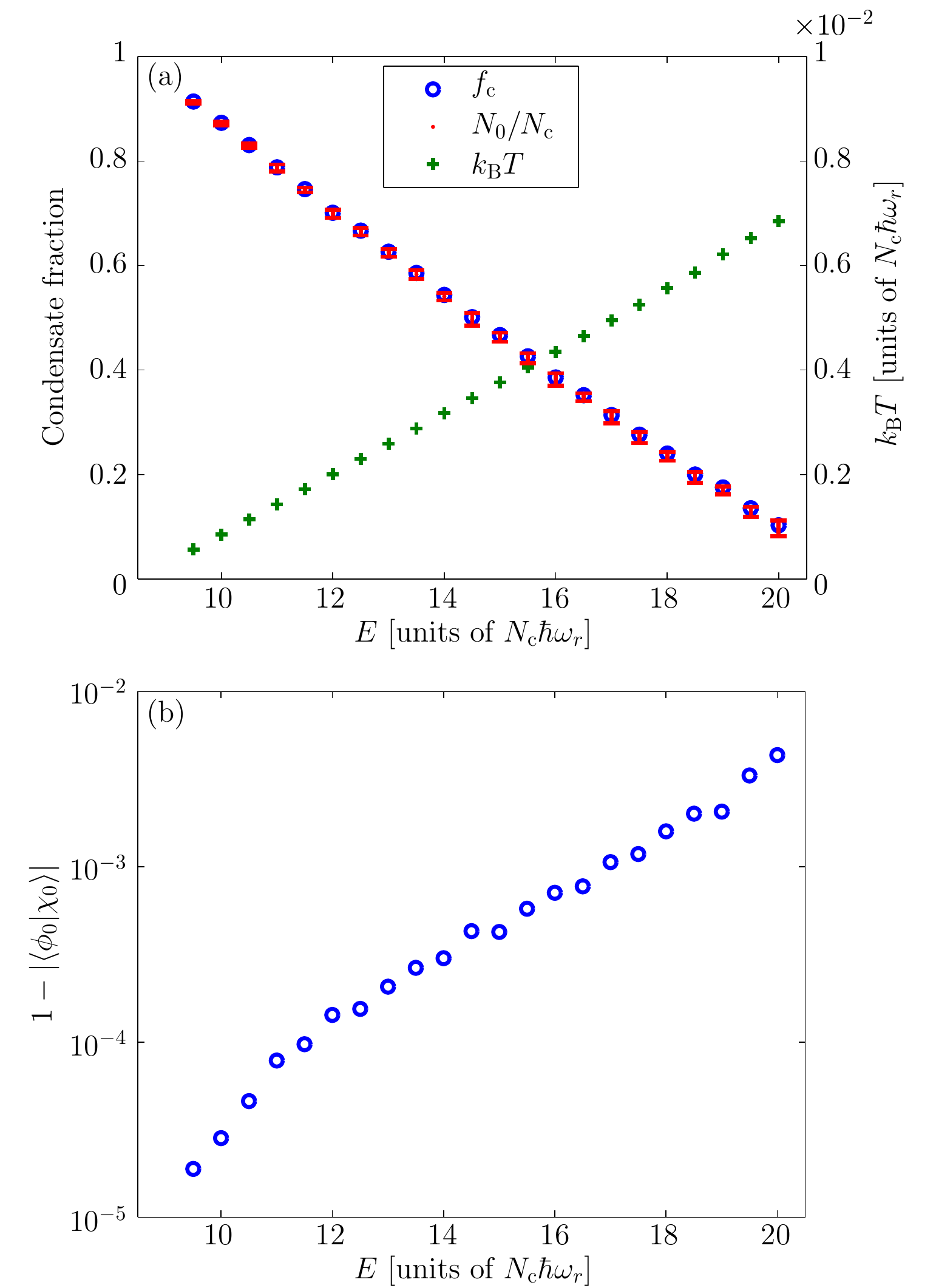}
	\caption{\label{fig:cfrac_mu} (Color online) (a) Condensed fraction of the classical field, as determined by the Penrose-Onsager procedure (circles) and by the time-averaging procedure (dots with error bars). Plusses indicate the microcanonical temperature of the field.  (b) Discrepancy $1-|\langle\phi_0|\chi_0\rangle|$ between the (unit-normalized) first moment $\phi_0(\mathbf{x})$ and the most highly occupied natural orbital $\chi_0(\mathbf{x})$ of the one-body density matrix.}
\end{figure}

We consider here the norm square $N_0$ of the first moment $\phi(\mathbf{x};\lambda_0)$ defined as in Sec.~\ref{subsec:first_moment}, for various values of the classical-field energy $E[\psi]$.  In Fig.~\ref{fig:cfrac_mu}(a) we present estimates $N_0$ for a range of classical-field energies, and compare them with the condensate occupations calculated by the PO approach.  
The corresponding classical-field temperatures, calculated using the Rugh methodology \cite{Rugh97,Davis03,Davis05}, are also included in the figure.  In practice, we calculated the PO condensate by constructing the one-body density matrix [Eq.~(\ref{eq:density_mtx})] from 3000 equally-spaced samples of the classical-field taken from a period of $1200\omega_r^{-1}$ of the field evolution.  We then divided this period into $30$ consecutive sub-periods of length $40\omega_r^{-1}$ which we sampled at a higher resolution (1000 samples per sub-period), from which we obtained $30$ separate estimates of the classical-field first moment $\phi(\mathbf{x};\lambda_0)$.  In each sub-period we obtain the mean field as the time-averaged field of maximal norm, and we obtain (generally) distinct estimates of $N_0$, $\phi_0(\mathbf{x})$ and $\lambda_0$ from each series.  In this way we exploit the ergodic character of the classical field to emulate sampling from an ensemble of similarly prepared trajectories (see Sec.~\ref{subsubsec:temporal_coherence}).  The red (gray) data points in Fig.~\ref{fig:cfrac_mu}(a) and their error bars represent in each case the mean and standard deviation of the norm squares of the $30$ estimates of the mean-field.  We observe that these estimates agree very closely with the PO condensate fractions $f_\mathrm{c}$ (blue circles) throughout the range of energies presented.  

We also compare the mean-field orbitals obtained from the time-averaging procedure with the condensate orbitals obtained from the PO approach (see Sec.~\ref{subsubsec:id_first_moment}). In Fig.~\ref{fig:cfrac_mu}(b) we plot the the quantity $1-|\langle \phi_0|\chi_0\rangle|$ (averaged over the 30 estimates) as a measure of the discrepancy between the two orbitals.  We observe that for the energies presented ($E\leq20N_\mathrm{c}\hbar\omega_r$) the mean discrepancy is $<10^{-2}$. At higher energies (corresponding to condensate fractions $f_\mathrm{c}<0.1$), our temporal-coherence approach to identifying the condensate begins to break down: the mean-field orbital $\phi_0(\mathbf{x})$ fails to match the PO orbital $\chi_0(\mathbf{x})$ (i.e. $|\langle \phi_0|\chi_0\rangle|<0.9$) in an increasing fraction of estimates as the condensate fraction $f_\mathrm{c}\rightarrow0$, and so for clarity we have not presented estimates for these energies here.  This point is discussed further in Sec.~\ref{subsubsec:phase_freq}. 
\subsubsection{Condensate frequency}\label{subsubsec:phase_freq}
\begin{figure}
	\includegraphics[width=0.45\textwidth]{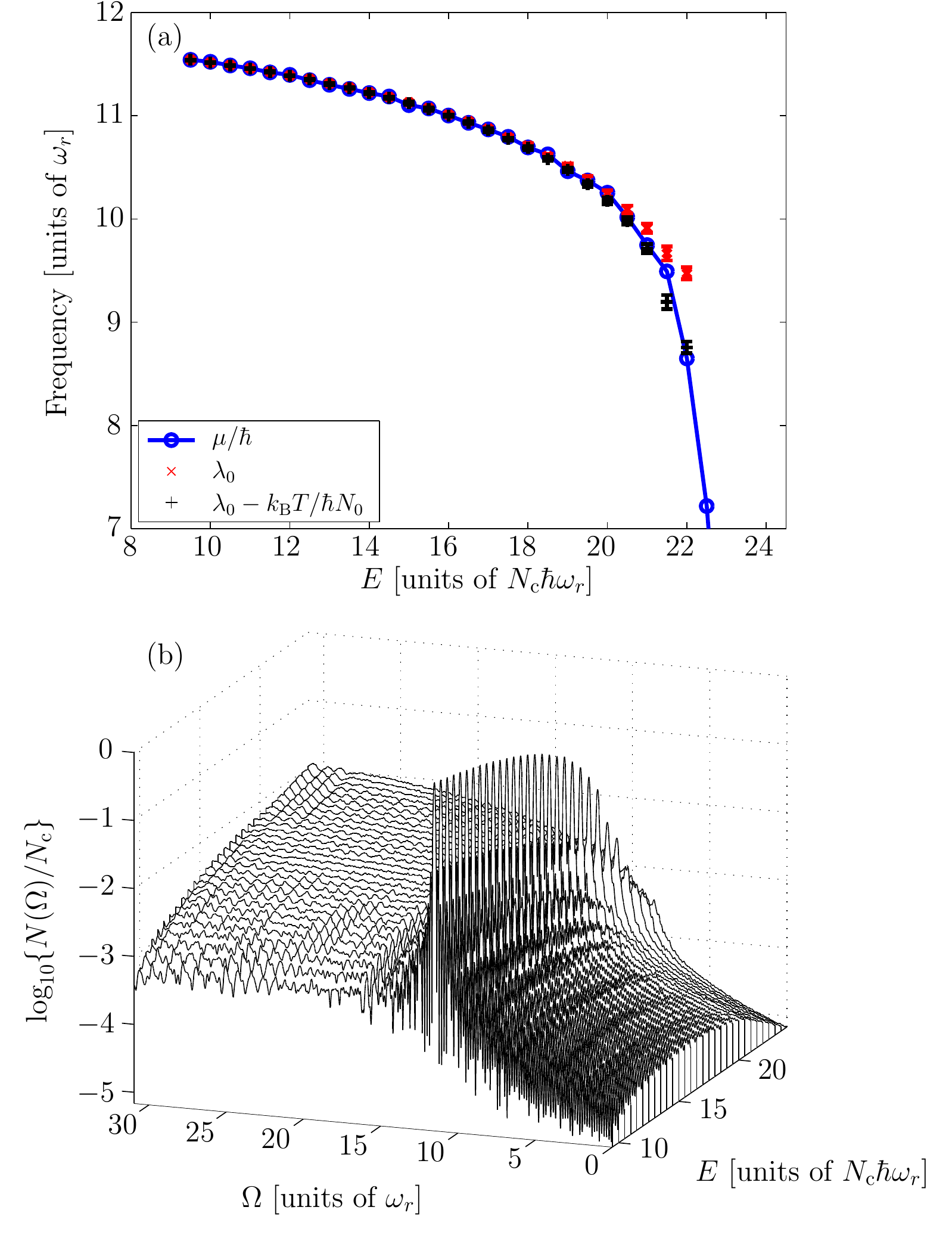}
	\caption{\label{fig:energy_frequency_carpet} (Color online) (a) Condensate frequency obtained from the time-averaging procedure (red crosses), and microcanonical chemical potential of the classical field (blue circles), for classical field equilibria with different energies. Black plusses plot the RHS of Eq.~(\ref{eq:mu_lambda_relation}). (b) Norm square of the first moment as a function of the phase-rotation frequency of the frame in which it is constructed (i.e. space-integrated power spectrum of the field), for classical-field simulations with different energies.}
\end{figure}

We consider here the dependence of the condensate frequency $\lambda_0$ on the energy of the classical field.  By our analogy between the first moment of the classical field and the condensate wavefunction in mean-field theories (Sec.~\ref{subsec:first_moment}), we associate this condensate frequency with the \emph{condensate eigenvalue} appearing in such theories, which is itself closely related to the thermodynamic chemical potential of the degenerate Bose-gas system \cite{Morgan00}.  In Fig.~\ref{fig:energy_frequency_carpet}(a) we plot estimates of the condensate frequency (red crosses), together with the thermodynamic chemical potential $\mu$ (blue circles with connecting line) of the classical field obtained from the Rugh analysis.  At the very highest energies, we present results only for ensemble members for which our first moment analysis and the PO approach agree (i.e. $\phi_0$ and $\chi_0$ overlap to within 10\%).  We observe that the condensate frequencies $\lambda_0$ and the chemical potentials $\mu$ agree very well for energies $E \lesssim 20N_\mathrm{c}\hbar\omega_r$.  Above this energy, the condensate frequencies $\lambda_0$ are consistently greater than the chemical potentials.  This is expected behavior, as at a fixed \emph{total} number of system particles, the two quantities differ by a factor of order $1/N_\mathrm{cond}$, where $N_\mathrm{cond}$ is the condensate occupation \cite{Proukakis98,Morgan00}.  Davis \emph{et al.} \cite{Davis02} argued that equipartition of energy in the classical-field model predicts the relationship  
\begin{equation}\label{eq:mu_lambda_relation}
	\mu = \hbar\lambda_0 - \frac{k_\mathrm{B}T}{N_0}.
\end{equation}
In Fig.~\ref{fig:energy_frequency_carpet}(a), we plot the quantity $\lambda_0 - k_\mathrm{B}T/\hbar N_0$ (black plusses), where the temperature $T$ is that obtained from the method of Rugh, and thus find that our results are in reasonable agreement with the prediction of Eq.~(\ref{eq:mu_lambda_relation}). 

We now consider how the total power spectrum of the classical field varies as a function of the field energy. In Fig.~\ref{fig:energy_frequency_carpet}(b) we plot the power spectrum $N(\Omega)$, averaged over the 30 individual $40\omega_r^{-1}$ sampling periods, for field energies in the range $E\in[9.5,24]N_\mathrm{c}\hbar\omega_r$.  At the lowest energies the behavior of $N(\Omega)$ is as in Fig.~\ref{fig:power_spectrum_in_trap}(b): the function exhibits a prominent peak which we identify with the condensate, and a broad background we associate with thermal excitations.  As the energy (and thus temperature) of the field is increased, the condensate peak decays, and the `wing' of thermal excitations grows until it is of the same magnitude as the condensate peak, and at the highest temperatures only the thermal background remains. This explains why our approach to identifying the condensate begins to fail as the temperature approaches the phase transition: although a temporally coherent condensate may still be present in the field, it becomes increasingly likely that the peak power in any particular estimate of the power spectrum corresponds instead to thermally occupied modes, which eventually swamp the condensate completely.
\section{Pairing correlations}\label{sec:pairing_correlations}
In the previous section, we have identified that the condensate present in the classical field can be well-characterized as the time-average of the appropriately frequency-shifted field.  We now show that more general anomalous moments of the field can be obtained from time-averages in the same phase-rotating frame.  In this approach, condensation in the classical field is thus accompanied by the appearance of anomalous moments of all orders, in direct analogy to the emergence of general anomalous correlation functions in symmetry-breaking accounts of Bose-Einstein condensation \cite{Griffin96,Proukakis96}.

In terms of the Fock-space decomposition $\hat{\Psi}(\mathbf{x}) = \sum_i \hat{a_i} Y_i(\mathbf{x})$, the emergence of a mean field in the second-quantized formalism is equivalent to the appearance of nonzero first moments $\{\langle \hat{a_i} \rangle \}$.  The next-simplest anomalous averages, the quadratic moments $\{\langle \hat{a_i} \hat{a_j} \rangle \}$ (and their conjugates), arise due to the effect of interactions which `mix' the single-particle creation and annihilation operators to form quasiparticle operators $\hat{b} \sim u \hat{a} + v^* \hat{a}^\dagger$ \cite{Blaizot86}.  Consequently, the occupation of quasiparticle modes results in the appearance of nonzero moments of single-particle operators of the form $\langle \hat{a_i} \hat{a_j} \rangle$, which represents correlations between \emph{pairs} of particles.  Like the mean field itself, these moments are formally zero in a state of fixed total particle number, although analogous quantities can be defined in particle-conserving terms \cite{Morgan00}.  Due to the appearance of these pairing correlations, in order to accurately characterize the weakly interacting Bose gas and its excitations, one must consider not only the one-body density matrix $\rho_{ij}=\langle \hat{a}_j^\dagger \hat{a}_i \rangle$, but also the \emph{pair matrix} $\kappa_{ij}=\langle \hat{a}_j\hat{a}_i \rangle$ \cite{Blaizot86}.

In the remainder of this section, we will demonstrate the application of our temporal averaging procedure to the evaluation of quadratic anomalous moments of the classical field: by estimating the pair matrix $\kappa(\mathbf{x},\mathbf{x}')=\langle \psi(\mathbf{x})\psi(\mathbf{x}')\rangle$ of the noncondensed component of the field, we calculate the \emph{anomalous density} which characterizes pairing correlations in the thermal component of the field. We note that signatures of such pairing correlations have been observed previously in classical-field calculations \cite{Wright09a}, where anomalous values $g^{(2)}_i=\langle |a_i|^4 \rangle/\langle |a_i|^2 \rangle^2 > 2$ were obtained for the second-order coherence functions of density-matrix eigenmodes.  We note also that a temporal signature of the anomalous density has previously been observed \cite{Brewczyk04} in homogeneous classical-field simulations, in which the anomalous density is uniform.
\subsection{Methodology}\label{subsec:methodology}
We seek here to characterize pairing correlations in the thermal component of the classical field, i.e., the component of the field \emph{orthogonal} to the condensate \cite{Hutchinson00}, which is obtained by projecting out the condensed component of $\psi(\mathbf{x},t)$, i.e. 
\begin{equation}\label{eq:project}
	\psi^\perp(\mathbf{x},t) = \psi(\mathbf{x},t) - \phi_0(\mathbf{x})\int d\mathbf{x}' \phi_0^*(\mathbf{x}')\psi(\mathbf{x}',t)d\mathbf{x}'.
\end{equation}
It is important to note that we form $\psi^\perp(\mathbf{x},t)$ on a given ($40\omega_r^{-1}$) time period by projecting out the mean field obtained over the \emph{same} period, so that (anomalous) averages constructed from $\psi^\perp(\mathbf{x},t)$ over this period are formed on the same footing as the mean field itself.  We transform $\psi^\perp(\mathbf{x},t)$ to the same phase-rotating frame as the condensate, forming $\tilde{\psi}^\perp(\mathbf{x},t)=e^{i\lambda_0t}\psi^\perp(\mathbf{x},t)$, and then calculate the pair matrix 
\begin{equation}\label{eq:pair_matrix}
	\kappa^\perp(\mathbf{x},\mathbf{x}') = \langle \tilde{\psi}^\perp(\mathbf{x})\tilde{\psi}^\perp(\mathbf{x}')\rangle_t.
\end{equation}
The most well-known characterization of the anomalous correlations described by the pair matrix is given by the \emph{anomalous density} \cite{Griffin96} which we identify as the diagonal part of the pair matrix
\begin{equation}\label{eq:anomalous_density}
	m(\mathbf{x}) = \langle \tilde{\psi}^\perp(\mathbf{x})\tilde{\psi}^\perp(\mathbf{x})\rangle_t \equiv \kappa^\perp(\mathbf{x},\mathbf{x}). 
\end{equation}
We find that the general form of $m(\mathbf{x})$ is apparent from a single estimate of $\kappa^\perp(\mathbf{x},\mathbf{x}')$, over a temporal period $40\omega_r^{-1}$.  However, large fluctuations are present in such a single estimate, which is to be expected, as the correlations we seek to resolve here are rather subtle as compared, for example, to the coherence of the condensate.  In order to resolve the anomalous density more clearly, we therefore average over multiple estimates of $m(\mathbf{x})$;  i.e., for each of $30$ consecutive $40\omega_r^{-1}$ periods, we form both the mean field $\phi_0(\mathbf{x})$ and the corresponding anomalous density $m(\mathbf{x})$.  The phase of $m(\mathbf{x})$ is only meaningful in relation to the phase of the mean field $\phi_0(\mathbf{x})$ itself, and so for convenience, we choose the overall phase of the classical field in each sampling period such that $\phi_0(\mathbf{x})$ is maximally real.  This choice of the phase of $m(\mathbf{x})$ relative to a real and positive condensate wavefunction corresponds to the traditional choice in mean-field theories.  Forming multiple estimates of $m(\mathbf{x})$ in this way allows us to calculate both the mean and the variance of this quantity, as we show in Sec.~\ref{subsubsec:anom_avg_results}.  
\subsection{Anomalous density}\label{subsubsec:anom_avg_results}
\begin{figure}
	\includegraphics[width=0.45\textwidth]{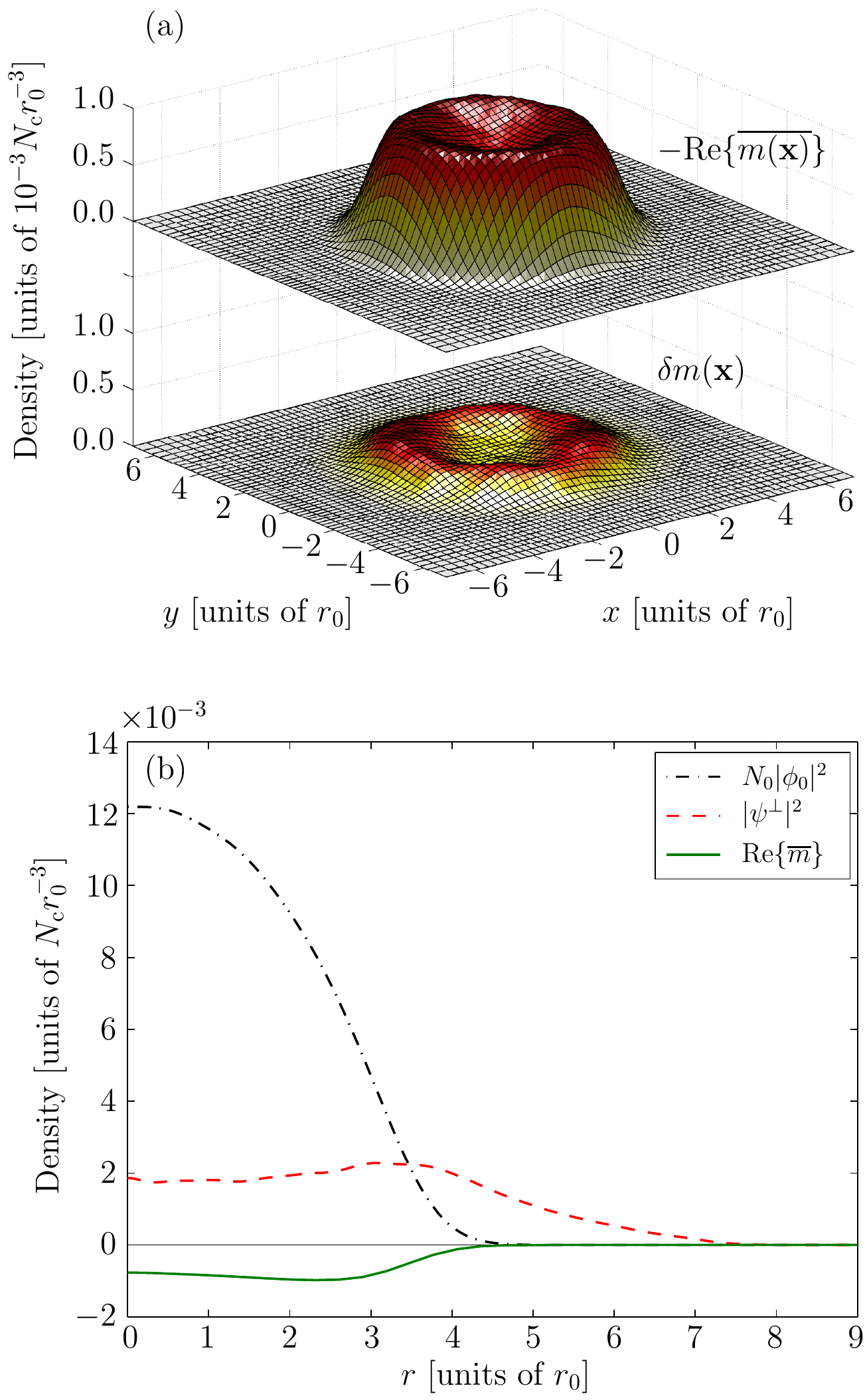}
	\caption{\label{fig:anomalous_average} (Color online) Anomalous density $m(\mathbf{x})$ of the field, for case $E=14.5N_\mathrm{c}\hbar\omega_r$. (a) Shape of $-\mathrm{Re}\{\overline{m(\mathbf{x})}\}$ on a slice through plane $z=0$ (upper surface), and standard deviation $\delta m(\mathbf{x})$ of anomalous-density estimates on the same plane (lower surface).  (b) Azimuthally averaged density of the condensate mode (as determined by the time-averaging), complementary (orthogonal) thermal component of the field, and anomalous density, in the plane $z=0$.}
\end{figure}

The mean anomalous density calculated by the procedure described in Sec.~\ref{subsec:methodology} is mostly real and negative (i.e., has phase opposite to that of the mean field), in agreement with the results of mean-field theory calculations \cite{Proukakis98,Hutchinson00,Bergeman00}, but exhibits some small complex-valued fluctuations due to the finite ensemble size.  Denoting averages over estimates by an overbar, we plot in Fig.~\ref{fig:anomalous_average}(a) the negative $-\mathrm{Re}\{\overline{m(\mathbf{x})}\}$ of the mean anomalous density on the $z=0$ plane, and the local standard deviation in estimates $\delta m(\mathbf{x}) = [\overline{|m(\mathbf{x})|^2} - |\overline{m(\mathbf{x})}|^2]^{1/2}$ of the anomalous density on this plane, calculated for a simulation with $E=14.5N_\mathrm{c}\hbar\omega_r$ (for which the condensate fraction $f_\mathrm{c}=0.50$).
The anomalous density has the spatial structure expected from mean-field calculations \cite{Proukakis98,Hutchinson00,Bergeman00}: it resides primarily in the region where the condensate exists, and its absolute value exhibits a shallow `dip' in the center of the trap.  The standard deviation $\delta m(\mathbf{x})$ indicates that the greatest variance in density estimates occurs around the (circular) maximum of $|m(\mathbf{x})|$, while much less variation occurs in estimates of the density in the central dip. 

The anomalous density shown here exhibits a very high degree of rotational symmetry about the $z$~axis, but in general the anomalous density we obtain is distorted (the central `dip' in its absolute value becomes saddle-shaped along some random axis).  We identify this as a result of persistent center-of-mass (dipole) excitations of the field \cite{Dobson94},  which are `frozen in' during the thermalization of the field.  More generally one might regard the classical field as having condensed into an excited center-of-mass mode, and consider the correlations of the field in a frame following this motion \cite{Pethick00}.  In Fig.~\ref{fig:anomalous_average}(b) we plot the azithumally averaged anomalous density on the plane $z=0$, together with the similarly averaged densities of the condensate [$N_0|\phi_0(\mathbf{x})|^2$] and the orthogonal thermal component of the field [$|\psi^\perp(\mathbf{x})|^2]$, for comparison.  We observe that the magnitude of the anomalous density in the center of the trap is an appreciable fraction of that of the (normal) thermal component of the field, in agreement with Refs.~\cite{Proukakis98,Hutchinson00,Bergeman00}.  
\subsection{Dependence on field energy}
\begin{figure}
	\includegraphics[width=0.45\textwidth]{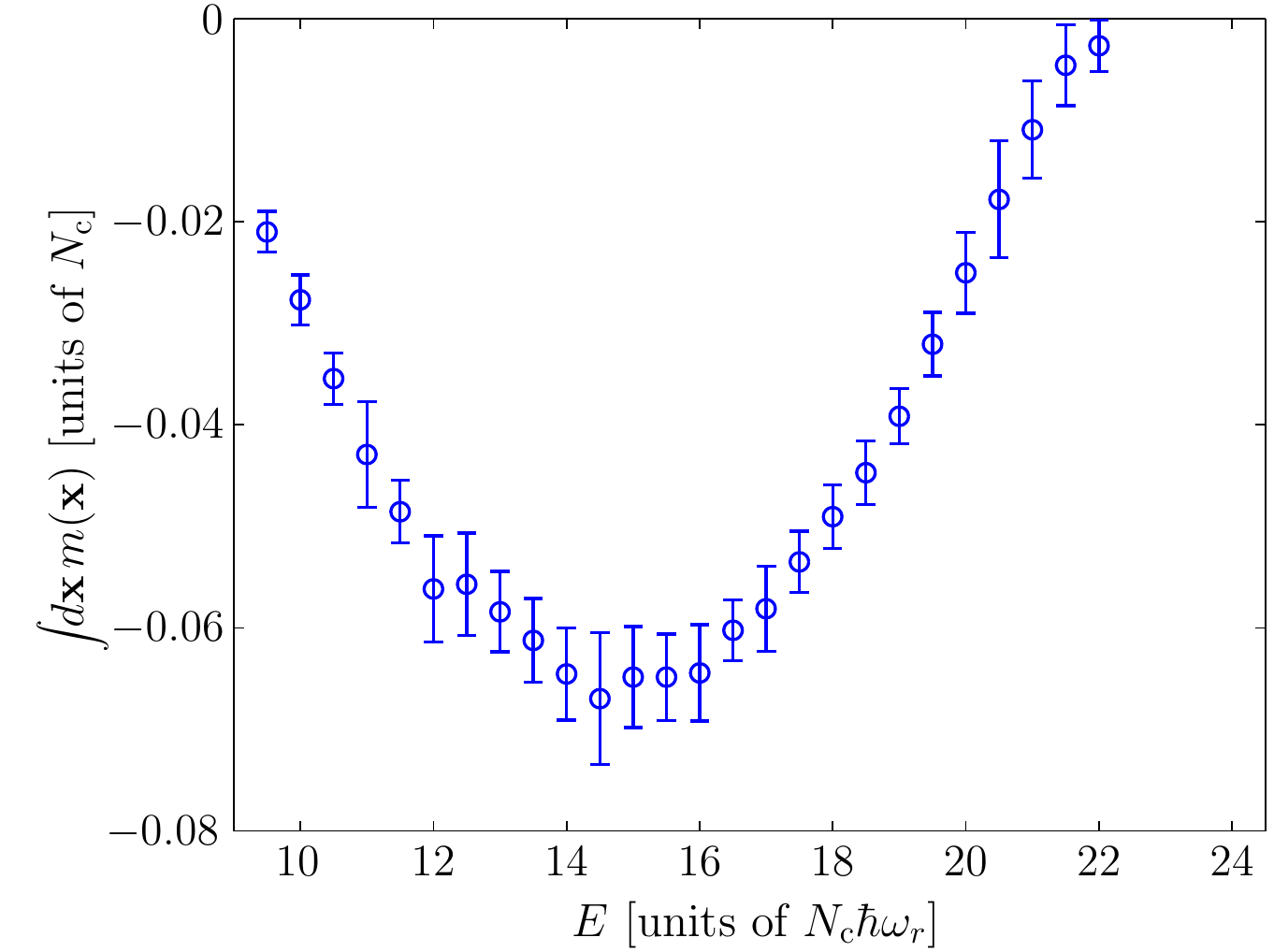}
	\caption{\label{fig:integrated_m} (Color online) Integrated value of the anomalous density $\int\!d\mathbf{x}\,m(\mathbf{x})$ as a function of field energy.}
\end{figure}

Finally, we consider the dependence of the anomalous density on the energy (or equivalently, the temperature) of the projected classical field.  In mean-field theories, the anomalous density (after any renormalization \cite{Proukakis98,Hutchinson98,Morgan00}) becomes small as the temperature of the system approaches zero (due to the weak occupation of quasiparticle modes in this limit), and also as it approaches the critical temperature (due to the quasiparticle modes becoming more single-particle-like as the condensate is depleted).  This behavior is often cited as a justification for the neglect of the anomalous density in self-consistent theories (the so-called Popov approximation \cite{Griffin96}) in these two limits.  In order to characterize more fully the temperature-dependent behavior of the anomalous density, we follow \cite{Hutchinson00} and calculate its integrated value $M\equiv \int \! d\mathbf{x}\, m(\mathbf{x})$.  In Fig.~\ref{fig:integrated_m} we plot the real part of $M$ (neglecting a small imaginary part that arises from incomplete convergence of the averaging -- see Sec.~\ref{subsubsec:anom_avg_results}) as a function of the classical-field energy.  
As in Sec.~\ref{subsec:energy_dependence}, we present for each energy the mean and standard deviation (error bars) of only those estimates obtained from averaging periods which produced an accurate condensate mode (Sec.~\ref{subsec:first_moment}).  We observe that the behavior of $M(E)$ is consistent with the results of mean-field theories \cite{Hutchinson00,Bergeman00}, with its absolute value $|M|$ reaching its maximum at intermediate energies (temperatures), and rapidly decreasing as we approach both the zero-temperature and critical regimes.

We note that the well-known issues of ultraviolet divergence of the anomalous density in mean-field theories arise from the \emph{zero-point} occupation of quasiparticle modes \cite{Proukakis98,Hutchinson98,Morgan00}, which is of course not present in the classical-field model.  Also, although the results of classical-field calculations are in general dependent on the cutoff energy,  the contribution to the anomalous density from successive quasiparticle modes rapidly decreases with increasing energy of the modes, as the modes return to a single-particle structure.  The requirement (for our treatment) that the anomalous density is well-contained in the low-energy region (condensate band \cite{Blakie08}) described by the PGPE is thus precisely the requirement that the cutoff is effected at such an energy that the interacting Hamiltonian has become approximately diagonal in the single particle basis $\{Y_k(\mathbf{x})\}$ (satisfied in practice for $E_R/\mu \gtrsim 3$ \cite{Gardiner97b,Gardiner03}).  

Finally, we note that the maximum (absolute) value of the integrated anomalous density occurs when $f_\mathrm{c}\approx0.5$, but remind the reader that this refers only to the proportion of the \emph{below-cutoff} field which is condensed.  Although the entire anomalous density should be well-described by the low-energy Hamiltonian PGPE dynamics, one would have to include the contribution of above-cutoff atoms to the \emph{normal} thermal density of the field \cite{Blakie08} in order to draw quantitative comparisons with (e.g.) the mean-field theory calculations of \cite{Hutchinson00}.
\section{Conclusions}\label{sec:conclusions}
We have demonstrated that in the Hamiltonian PGPE theory, classical-wave condensation is accompanied by long-range temporal coherence limited only by the slow diffusion of the condensate phase.  This gives rise to the appearance of a nonzero first moment of the field, as defined by short-time averages in an appropriate phase-rotating frame.  We identified the emergence of this moment with the concept of $\mathrm{U}(1)$-symmetry breaking that is central to self-consistent mean-field theories.  We showed that the mean field obtained by short-time averaging agrees well with the condensate identified by the standard Penrose-Onsager approach, except for close to the critical regime associated with the transition to the normal phase.  The condensate eigenfrequency obtained by this temporal analysis exhibits the behavior predicted for the condensate eigenvalue in the most sophisticated mean-field approaches \cite{Morgan00}, i.e., it agrees closely with the thermodynamic chemical potential at low energies and diverges away from it in inverse proportion to the condensate occupation as the phase transition is approached.  By calculating the pair matrix and anomalous density of the noncondensed component of the field, we demonstrated explicitly that time averages in the frame rotating at the condensate frequency allow the calculation of more general anomalous moments.  We observed the anomalous density to exhibit the expected behavior \cite{Hutchinson00}, with its magnitude reaching its maximum at intermediate temperatures and decreasing as both the $T=0$ and critical regimes are approached.  
\begin{acknowledgments}
We wish to acknowledge discussions with A. S. Bradley, M. J. Davis, and C. W. Gardiner.  
This work was supported by the New Zealand Foundation for Research, Science and Technology under Contract No. NERF-UOOX0703.  TMW acknowledges financial support from the Australian Research Council Centre of Excellence for Quantum-Atom Optics (CE0348178).
\end{acknowledgments}
\appendix
\section{Phase diffusion of the condensate}\label{app:AppendixA}
\subsection{Single phase-diffusive mode}
Let us first consider a single-mode model of the condensate, in which the amplitude $a_0(t)$ of the condensate mode (with condensate frequency $\lambda_0$) exhibits phase diffusion.  We assume that the mode does not exhibit any number fluctuations, which is precisely the condition that $g_0^{(2)}=\langle |a_0|^4 \rangle_t / \langle |a_0|^2 \rangle^2_t = 1$, which is well satisfied away from the critical regime \cite{Bezett09b,Wright09a}.  We thus have $a_0=|a_0|e^{i\theta(t)}$, and defining $\varphi(t) \equiv \theta(t) - \theta(0)$, we assume $\mathrm{var}\{\varphi(t)\} \equiv \langle \varphi(t)^2\rangle - \langle \varphi(t) \rangle^2 = 2\gamma t$ \cite{Sinatra08}, where $\gamma$ is the (phase) \emph{diffusion coefficient}, and $\langle \cdots \rangle$ denotes an average over \emph{realizations} of the amplitude $a_0(t)$ (i.e., an ensemble average).   This is precisely the behavior of the \emph{Kubo oscillator} \cite{Gardiner04} stochastic process, which obeys the (Ito) stochastic differential equation 
\begin{equation}\label{eq:kubo_oscillator}
	da_0 = [(-i\lambda_0 - \gamma)dt + i\sqrt{2\gamma}dW(t)]a_0(t),
\end{equation}
where $dW(t)$ is a \emph{real} Wiener increment, which satisfies $\langle dW(t) dW(t') \rangle = \delta(t-t')dt$.  By studying this simple model we hope to gain insight into the behavior of our diffusive condensate mode.

We consider the power spectrum of the mode obtained over a period $T$,
\begin{equation}
	N^{(0)}(\Omega;T) = \Big|\frac{1}{T}\int_0^Tdt\;e^{i\Omega t} a_0(t) \Big|^2.
\end{equation}
This power spectrum is itself a stochastic process (developing in $T$), i.e., it varies between realizations of the oscillator.  We therefore consider its mean $\langle N^{(0)}(\Omega;T)\rangle$.  Using the known result $\langle a_0(t) a_0^*(s) \rangle = |a_0|^2\exp[-i\lambda_0(t-s) -\gamma|t-s|]$ \cite{Gardiner04}, we find 
\begin{eqnarray}\label{eq:N0_mean_general}
	\langle N^{(0)}(\Omega;T) \rangle &=& \frac{1}{T^2(\gamma^2+\Delta^2)^2} \Bigg\{ \gamma T (\gamma^2 + \Delta^2) \\
&&+\Big[e^{-\gamma T}\cos(\Delta\;T) - 1\Big](\gamma^2 - \Delta^2) - 2\gamma \Delta \sin(\Delta\;T)  \Bigg\}, \nonumber
\end{eqnarray}
which we have written in terms of $\Delta\equiv\Omega-\lambda_0$ for compactness.  In the limit of no diffusion ($\gamma \rightarrow 0$) we regain the result $N^{(0)}(\Omega;T) = |a_0|^2 \mathrm{sinc}^2\Big[\frac{1}{2}(\Omega-\lambda_0)T\Big]$ appropriate to the resolution of a single frequency by a measurement of finite duration $T$.  In the limit of a measurement made on a time scale long compared with the characteristic diffusion time, i.e. $\gamma T \gg 1$, we regain the Lorentzian spectrum of the Kubo oscillator $N^{(0)}(\Omega;T)=(2|a_0|^2\gamma/T)/[\gamma^2+(\Omega-\lambda_0)^2]$.  From Eq.~(\ref{eq:N0_mean_general}), the power measured at the underlying frequency $\lambda_0$ of the oscillator can be obtained by setting $\Delta=0$, giving 
\begin{equation}\label{eq:mean_decay}
	\langle N^{(0)}(\lambda_0;T) \rangle = |a_0|^2\frac{2}{(\gamma T)^2} \Big[\gamma T - (1 - e^{-\gamma T})\Big].
\end{equation}
For short time periods $T \ll 1/\gamma$ (such as we consider in the main text), the norm square of the mean field decays like $\sim 1 - \gamma T/3$, while at long times it decays like $\sim 2/\gamma T$.   It is important to note that this same functional form would be exhibited by (e.g.) a complex Ornstein-Uhlenbeck process \cite{Gardiner04}, which one might reasonably assume as a model for a thermally occupied mode \cite{Gardiner00,Stoof01,Sinatra07} in a classical-field approximation:  the `bare' (i.e. infinite sampling time) power spectrum of such a mode is similarly Lorentzian, and so we expect the same behavior both for two-time correlations [$|\langle a^*(t) a(0) \rangle| \sim e^{-\gamma t}$] and for the measured power $N(\Omega;T)$, and the two cases (i.e. condensate and thermal mode) are thus distinguished only by the \emph{time scales} on which the power decays.  \emph{Qualitative} differences between the two types of mode thus only appear in second-order (and higher) correlation functions, which are sensitive to \emph{number} fluctuations.

\subsection{Multimode description}
In general the condensate mode is only one mode in a multimode field which contains other, thermally occupied modes.  We expect the thermal field to be well described in the basis of Bogoliubov modes $\{(u_i,v_i)\}$ orthogonal to the condensate mode \cite{Castin01}, and thus assume
\begin{equation}
	\psi(\mathbf{x},t) = a_0(t)\chi_0(\mathbf{x}) + \sum_j \Big( b_j(t) u_j(\mathbf{x}) + b_j^*(t)  v_j^*(\mathbf{x}) \Big),
\end{equation}
where, to gain simple insight into our measurements of the field, we assume that the $\{b_j(t)\}$ are complex Ornstein-Uhlenbeck processes which are uncorrelated with one another and with the condensate.  The total power spectrum of the field is thus 
\begin{eqnarray}
	\langle N(\Omega;T) \rangle &=& \langle N^{(0)}(\Omega;T)\rangle \int d\mathbf{x} |\chi_0(\mathbf{x})|^2   \\
	&&+ \sum_j \langle N^{(j)}(\Omega;T) \rangle \int d\mathbf{x} |u_j(\mathbf{x})|^2 + |v_j(\mathbf{x})|^2, \nonumber
\end{eqnarray}
where 
\begin{eqnarray}
	\langle N^{(j)}(\Omega;T) \rangle = \Big|\frac{1}{T}\int_0^T dt e^{i\Omega t} b_j(t) \Big|^2,
\end{eqnarray}
behave similarly to $\langle N^{(0)}(\Omega;T) \rangle$, except that they are centered on the frequencies $\epsilon^B_j/\hbar$ of the Bogoliubov modes, and attenuate much more rapidly with $T$ ($\gamma_j \gg \gamma_0$).  There is therefore power in the field at a range of frequencies, however, on times $T\gg 1/\gamma_j$ we have $\langle N(\Omega;T) \rangle \approx N^{(0)}(\Omega;T) \int d\mathbf{x} |\chi_0(\mathbf{x})|^2$ and, moreover, 
\begin{equation}
	\Big\langle \frac{1}{T} \int dt\, e^{i\lambda_0 t} \psi(\mathbf{x},t) \Big\rangle = \frac{1-e^{-\gamma T}}{\gamma T} |a_0|\langle e^{i\theta(0)}\rangle \chi_0(\mathbf{x}),
\end{equation} 
where the appearance of the expectation $\langle e^{i\theta(0)} \rangle$ of the initial complex phase emphasizes that the condensate phase varies randomly between ensemble members, `breaking' the $\mathrm{U}(1)$ symmetry in any particular realization.
\bibliographystyle{prsty}

\begin{thebibliography}{10}

\bibitem{Griffin96}
A. Griffin, Phys. Rev. B {\bf 53},  9341  (1996).

\bibitem{Burnett99}
K. Burnett,  in {\em Bose-Einstein Condensation in Atomic Gases}, Proceedings of the International School of Physics
"Enrico Fermi", Course CXL, edited by M. Inguscio, S. Stringari, and C.~E. Wieman (IOS Press, Amsterdam, 1999).

\bibitem{Hutchinson00}
D.~A.~W. Hutchinson {\it et~al.}, Journal of Physics B: Atomic, Molecular and
  Optical Physics {\bf 33},  3825  (2000).

\bibitem{Leggett01}
A.~J. Leggett, Rev. Mod. Phys. {\bf 73},  307  (2001).

\bibitem{Proukakis08}
N.~P. Proukakis and B. Jackson, Journal of Physics B: Atomic, Molecular and
  Optical Physics {\bf 41},  203002  (2008).

\bibitem{Gardiner97a}
C.~W. Gardiner, Phys. Rev. A {\bf 56},  1414  (1997).

\bibitem{Castin98}
Y. Castin and R. Dum, Phys. Rev. A {\bf 57},  3008  (1998).

\bibitem{Morgan00}
S.~A. Morgan, J. Phys. B {\bf 33},  3847  (2000).

\bibitem{Gardiner07}
S.~A. Gardiner and S.~A. Morgan, Phys. Rev. A {\bf 75},  043621  (2007).

\bibitem{Sinatra01}
A. Sinatra, C. Lobo, and Y. Castin, Phys. Rev. Lett. {\bf 87},  210404  (2001).

\bibitem{Blakie08}
P.~B. Blakie, A.~S. Bradley, M.~J. Davis, R.~J. Ballagh and C.~W. Gardiner, Advances In Physics {\bf 57},  363  (2008).

\bibitem{Goral01}
K. Goral, M. Gajda, and K. Rzazewski, Opt. Express {\bf 8},  92  (2001).

\bibitem{Norrie06a}
A.~A. Norrie, R.~J. Ballagh, and C.~W. Gardiner, Physical Review A (Atomic,
  Molecular, and Optical Physics) {\bf 73},  043617  (2006).

\bibitem{Polkovnikov03b}
A. Polkovnikov, Phys. Rev. A {\bf 68},  033609  (2003).

\bibitem{Blakie05}
{P.~B.~Blakie} and {M.~J.~Davis}, Phys. Rev. A {\bf 72},  063608  (2005).

\bibitem{Wright08}
T.~M. Wright, R.~J. Ballagh, A.~S. Bradley, P.~B.~Blakie and C.~W. Gardiner, Physical Review A (Atomic, Molecular, and Optical
  Physics) {\bf 78},  063601  (2008).

\bibitem{Penrose56}
{O. Penrose} and {L. Onsager}, Phys. Rev. {\bf 104},  576  (1956).

\bibitem{Sinatra08}
A. Sinatra and Y. Castin, Physical Review A (Atomic, Molecular, and Optical
  Physics) {\bf 78},  053615  (2008).

\bibitem{Lebowitz73}
J.~L. Lebowitz and O. Penrose, Physics Today {\bf 26},  23  (1973).

\bibitem{Davis01}
{M. J. Davis}, {S. A. Morgan}, and {K. Burnett}, Phys. Rev. Lett. {\bf 87},
  160402  (2001).

\bibitem{Rugh97}
H.~H. Rugh, Phys. Rev. Lett. {\bf 78},  772  (1997).

\bibitem{Bradley08}
A.~S. Bradley, C.~W. Gardiner, and M.~J. Davis, Phys. Rev. A {\bf 77},  033616
  (2008).

\bibitem{Wright09a}
T.~M. Wright, A.~S. Bradley, and R.~J. Ballagh, Physical Review A (Atomic,
  Molecular, and Optical Physics) {\bf 80},  053624  (2009).

\bibitem{Wright10a}
T.~M. Wright, A.~S. Bradley, and R.~J. Ballagh, Phys. Rev. A {\bf 81},  013610
  (2010).

\bibitem{Blaizot86}
J.-P. Blaizot and G. Ripka, {\em Quantum Theory of Finite Systems} (MIT Press,
  Cambridge, Massachusetts, 1986).

\bibitem{Graham02}
R. Graham, Nonlinear Phenom. Complex Syst. (Dordrecht, Neth.) {\bf 5},  349
  (2002).

\bibitem{NoteA}
{Note that the cutoff energy refers only to the \emph{single-particle} modes
  which the field is expanded on. The energies of excitations in the
  interacting field are raised due to the mean-field potential they experience
  (see also discussion in \cite{Wright08}).}

\bibitem{Fetter99}
A.~L. Fetter,  in {\em Bose-Einstein Condensation in Atomic Gases}, Proceedings of the International School of Physics
"Enrico Fermi", Course CXL, edited by M. Inguscio, S. Stringari, and C.~E. Wieman (IOS Press, Amsterdam, 1999).

\bibitem{Mora03}
C. Mora and Y. Castin, Phys. Rev. A {\bf 67},  053615  (2003).

\bibitem{NoteB}
{The identification of the condensate and its eigenfrequency in this manner
  bears some resemblance to the method of Feit \emph{et al.} \cite{Feit82} for
  the identification of Schr\"odinger eigenstates and eigenvalues from a
  spectral analysis of trajectories of the time-dependent Schr\"odinger
  equation. In contrast to that work, however, the condensate `eigenmode' we
  consider here emerges from the trajectories of a \emph{nonlinear} equation of
  motion, and can only be obtained from an analysis of the real-time field
  trajectories [or some other sampling of the PGPE microcanonical density
  Eq.~(\ref{eq:mu_density})].}

\bibitem{NoteC}
{We calculated (and fitted to) estimates of the powers at $T=4,8,\cdots,400
  \omega_r^{-1}$, but for clarity indicate only every tenth such estimate in
  Fig.~\ref{fig:decay}.}

\bibitem{Davis03}
M.~J. Davis and S.~A. Morgan, Phys. Rev. A {\bf 68},  053615  (2003).

\bibitem{Davis05}
M.~J. Davis and P.~B. Blakie, Journal of Physics A: Mathematical and General
  {\bf 38},  10259  (2005).

\bibitem{Proukakis98}
N.~P. Proukakis, S.~A. Morgan, S. Choi, and K. Burnett, Phys. Rev. A {\bf 58},
  2435  (1998).

\bibitem{Davis02}
M.~J. Davis, S.~A. Morgan, and K. Burnett, Phys. Rev. A {\bf 66},  053618
  (2002).

\bibitem{Proukakis96}
N.~P. Proukakis and K. Burnett, J. Res. Natl. Inst. Stand. Technol. {\bf 101},
  457  (1996).

\bibitem{Brewczyk04}
M. Brewczyk, P. Borowski, M. Gajda, and K. Rz{\c a}\.zewski, J. Phys. B {\bf
  37},  2725  (2004).

\bibitem{Bergeman00}
T. Bergeman, D.~L. Feder, N.~L. Balazs, and B.~I. Schneider, Phys. Rev. A {\bf
  61},  063605  (2000).

\bibitem{Dobson94}
J.~F. Dobson, Phys. Rev. Lett. {\bf 73},  2244  (1994).

\bibitem{Pethick00}
C.~J. Pethick and L.~P. Pitaevskii, Phys. Rev. A {\bf 62},  033609  (2000).

\bibitem{Hutchinson98}
D.~A.~W. Hutchinson, R.~J. Dodd, and K. Burnett, Phys. Rev. Lett. {\bf 81},
  2198  (1998).

\bibitem{Gardiner97b}
C.~W. Gardiner and P. Zoller, Phys. Rev. A {\bf 58},  536  (1998).

\bibitem{Gardiner03}
{C.~W.~Gardiner} and {M.~J.~Davis}, J. Phys. B {\bf 36},  4731  (2003).

\bibitem{Bezett09b}
A. Bezett and P.~B. Blakie, Physical Review A (Atomic, Molecular, and Optical
  Physics) {\bf 79},  033611  (2009).

\bibitem{Gardiner04}
C.~W. Gardiner, {\em Handbook of Stochastic Methods}, $3^\mathrm{rd}$ ed.
  (Springer-Verlag, Berlin, 2004).

\bibitem{Gardiner00}
C.~W. Gardiner and P. Zoller, {\em Quantum Noise}, $2^\mathrm{nd}$ ed.
  (Springer-Verlag, Berlin, 2000).

\bibitem{Stoof01}
H.~T.~C. Stoof,  in {\em Coherent Atomic Matter Waves}, Proceedings of the Les Houches Summer School
  of Theoretical Physics, Session LXXII, edited by R. Kaiser, C. Westbrook, and F. David
  (EDP Sciences \& Springer-Verlag, 2001). 

\bibitem{Sinatra07}
A. Sinatra, Y. Castin, and E. Witkowska, Physical Review A (Atomic, Molecular,
  and Optical Physics) {\bf 75},  033616  (2007).

\bibitem{Castin01}
Y. Castin,  in {\em Coherent Atomic Matter Waves}, Proceedings of the Les Houches Summer School
  of Theoretical Physics, Session LXXII, edited by R. Kaiser, C. Westbrook, and F. David
  (EDP Sciences \& Springer-Verlag, 2001). 

\bibitem{Feit82}
M.~D. Feit, J.~A. Fleck, Jr., and A. Steiger, J. Comput. Phys. {\bf 47},  412
  (1982).

\end{thebibliography}

\end{document}